\documentclass[12pt]{iopart}

\usepackage{iopams}
\usepackage{amssymb}
\usepackage{graphicx}
\usepackage{afterpage}
\usepackage{amsfonts}
\usepackage{braket}
\usepackage{textcomp}
\usepackage{xcolor}
\usepackage{soul}

\begin{document}

\title[]{Controlled coherent-coupling and dynamics of exciton complexes in a MoSe$_{\boldsymbol2}$ monolayer}

\author{Aleksander~Rodek$^1$, Thilo~Hahn$^2$, James~Howarth$^3$, Takashi~Taniguchi$^4$, Kenji~Watanabe$^5$, Marek~Potemski$^{1,6,7}$, Piotr~Kossacki$^1$, Daniel~Wigger$^8$, and Jacek~Kasprzak$^{1,9,10}$}

\address{$^1$Faculty of Physics, University of Warsaw, ul. Pasteura 5, 02-093 Warszawa, Poland}
\address{$^2$Institute of Solid State Theory, University of M\"unster, 48149 M\"unster, Germany}
\address{$^3$National Graphene Institute, University of Manchester, Booth St E, M13 9PL United Kingdom}
\address{$^4$International Center for Materials Nanoarchitectonics, National Institute for Materials Science,  1-1 Namiki, Tsukuba 305-0044, Japan}
\address{$^5$Research Center for Functional Materials, National Institute for Materials Science, 1-1 Namiki, Tsukuba 305-0044, Japan}
\address{$^6$Laboratoire National des Champs Magn\'{e}tiques Intenses, CNRS-UGA-UPS-INSA-EMFL, 25 Av. des Martyrs, 38042 Grenoble, France}
\address{$^7$CENTERA Labs, Institute of High Pressure Physics, PAS, 01- 142 Warsaw, Poland}
\address{$^8$School of Physics, Trinity College Dublin, Dublin 2, Ireland}
\address{$^9$Univ. Grenoble Alpes, CNRS, Grenoble INP, Institut N\'{e}el, 38000 Grenoble, France}
\address{$^{10}$Walter Schottky Institut and TUM School of Natural Sciences, Technische Universit\"at M\"unchen, 85748 Garching, Germany}

\ead{aleksander.rodek@fuw.edu.pl, piotr.kossacki@fuw.edu.pl, jacek.kasprzak@neel.cnrs.fr}

\vspace{10pt}
\begin{indented}
\item[]January 2023
\end{indented}

\begin{abstract}
Quantifying and controlling the coherent dynamics and couplings of
optically active excitations in solids is of paramount importance in
fundamental research in condensed matter optics and for their
prospective optoelectronic applications in quantum technologies.
Here, we perform ultrafast coherent nonlinear spectroscopy of a
charge-tunable MoSe$_2$ monolayer. The experiments show that the
homogeneous and inhomogeneous line width and the population decay of
exciton complexes hosted by this material can be directly tuned by
an applied gate bias, which governs the Fermi level and therefore
the free carrier density. By performing two-dimensional
spectroscopy, we also show that the same bias-tuning approach
permits us to control the coherent coupling strength between charged
and neutral exciton complexes.
\end{abstract}

%
%
%
%
%

\section{Introduction}
The ability to isolate monolayer flakes of semiconducting
transition-metal dichalcogenides (TMDs), such as MoSe$_2$, and the
discovery of their enhanced light-matter interaction at the
monolayer limit a decade ago~\cite{SplendianiNanoLett10, MakPRL10}
established a novel benchmark for semiconductor optics. The
combination of heavy effective masses and spin-valley
locking -- both inherent in the monolayer's band structure -- with
a large out-of-plane dielectric contrast result in particularly
strong Coulomb interactions among free carriers hosted by TMD
monolayers~\cite{RajaNatCom17, DinhVanTuanPRB18}. For the same
reason, the exciton transitions display a non-hydrogenic excitation
spectrum~\cite{ChernikovPRL14} and have large binding energies and
oscillator strengths~\cite{WangRMP18, StepanovPRL21}, such that
they dominate the optical response of TMD materials even at room temperature.

Further to that, the ability to stack different sorts of layered van
der Waals materials, such as TMDs, hexagonal boron nitride (hBN),
graphene or graphite, into van der Waals
heterostructures~\cite{GeimNature13, CadizPRX17}, facilitates the
fabrication of optoelectronic devices~\cite{MakNatNano12,
ZengNatNano12, MakScience14, KioseoglouAPL12, CaoNatCommun12}.
Sandwiching layered TMDs or other materials between high quality hBN
diminishes the electronic disorder and protects them from
degradation by ambient agents. In general, the hBN-encapsulation
procedure reduces the exciton's spectral inhomogeneous broadening
$\hbar\sigma$, thus improving the optical performance of
TMDs~\cite{WierzbowskiSciRep17, Ayayi2DMater17, MancaNatCom17,
LindlauNatCom2018}. It even enables reaching the homogeneous limit
for the exciton's spectral line shape~\cite{JakubczykACSNano19,
BoulePRM20}. This has been a milestone for monolayer spectroscopy, as
it enhances the visibility of phenomena previously blurred by
different types of inhomogeneity, with the appearance of a novel
destructive photon echo being an example~\cite{HahnNJP21}.

We are employing a four-wave mixing (FWM) spectroscopy approach to study the coherent optical properties of TMD exciton complexes. This method is recently gaining more popularity for the investigation of layered semiconductors: Besides the fundamental studies of the exciton coherence and population dynamics in different TMD systems~\cite{JakubczykNanoLett16, Jakubczyk2DMat17, JakubczykACSNano19} it was also used to probe intervalley scattering processes~\cite{hao2016direct}, valley coherence times~\cite{Hao2DMat17}, and mechanisms of exciton broadening~\cite{MoodyNatCom15}. In particular it allowed for first observations of biexcitons~\cite{conway2022direct, HaoNatCom17} or coupling between different exciton states~\cite{HaoNatCom17} in various TMD materials.

Electron beam lithography and metal deposition can further be employed
complement van der Waals heterostructures with electronic contacts and gates~\cite{RossNatCom13, JonesNatCom13}.
In this way, one can inject carriers into TMD monolayers, permitting
us to control the free electron density $n_{\rm e}$ in our sample. This is crucial,
not only for optoelectronic applications, but also when
exploring the fundamental physics of correlated many-body systems in
solids; landmark examples being the recent revelation of Wigner
crystals in a gate-tunable MoSe$_2$ monolayer~\cite{SmolenskiNature21} or
optical sensing of the quantum Hall effect in graphene~\cite{PopertNanoLett22}

The presence of free carriers modifies the excitonic optical
response substantially~\cite{EfimkinPRB17,ChangPRB19}, similarly as
was shown and partially explained for semiconductor quantum
wells~\cite{KossackiPRB99,KossackiPRB05,HawrylakPRB91,BrinkmannPRB99}.
Firstly, it is supposed that free carriers affect the relative
absorption (oscillator or effective dipole strength) and
the possible couplings between neutral excitons (X) and (negatively) charged
excitons (X$^-$), also called trions. Secondly, the exciton's
dephasing and therefore the spectral line shape should be sensitive
to the electron density $n_{\rm e}$~\cite{KochPRB95}, in an analogous way as it depends on
the total exciton density, through the mechanisms called excitation
induced dephasing~\cite{SchultheisPRL86, WangPRL93, BoulePRM20},
sometimes described by a local field effect~\cite{WegenerPRA90,
RodekNanophot21}. Thirdly, with increasing $n_{\rm e}$ the exciton
complexes are screened more efficiently from
intrinsic disorder, which in turn should reduce their inhomogeneous
broadening~\cite{RajaNatCom17, AkbariNanoLett22}, further
impacting the excitons' radiative decay rates~\cite{SavonaPRB06}. Here, we
address the basic interplay between free carriers $n_{\rm e}$,
excitons X, and trions X$^-$ by performing coherent nonlinear
spectroscopy~\cite{RossiRMP02,ShahSpringer99} of a gated and
hBN-encapsulated MoSe$_2$ monolayer.

\section{Sample \& Experiment}
\begin{figure}
    \centering
    \includegraphics[width=\columnwidth]{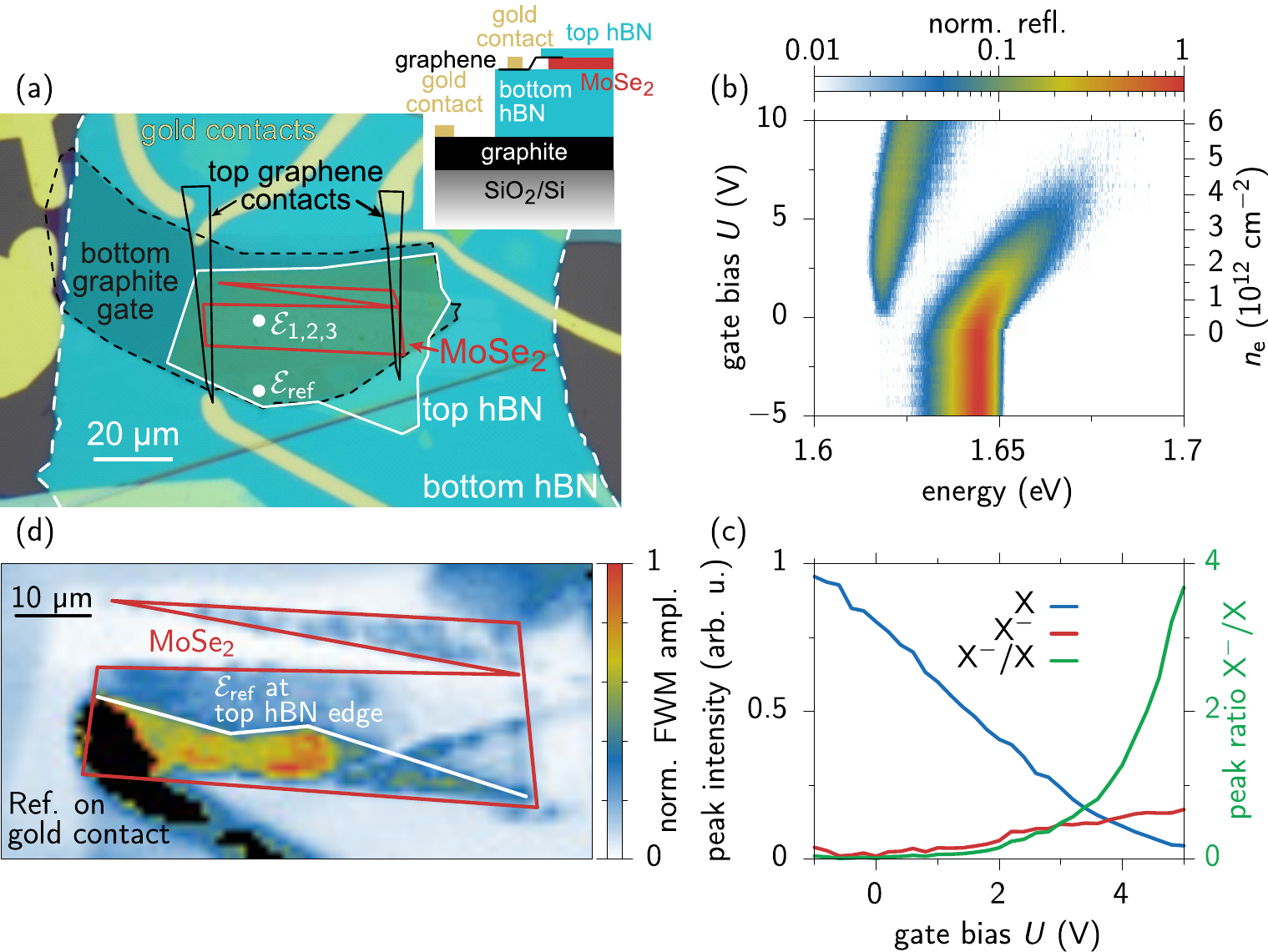}
    \caption{\textbf{Investigated MoSe$_{\boldsymbol 2}$ device and optical characterization.} (a) Optical
image of the heterostructure. The individual components are marked by colored lines and the white dots mark the positions of the laser beams for the FWM measurements and the inset shows the layer structure of the sample.
    (b) Reflectance contrast spectrum of the sample as a function of the
applied voltage showing the neutral (X) and negatively-charged
(X$^-$) exciton transitions. (c) Individual peak intensities X in
blue and X$^-$ in red from (b) and their ratio in green. (d) Imaging
of the time-integrated FWM at $\tau_{12}=1.2$\,ps. The black area in
the lower-left corner of the image is an experimental artefact
generated when the reference beam impinges on the highly reflecting
metallic contact $\sim10$\textmu m below the edge of the MoSe$_2$
monolayer. The outline of the MoSe$_2$ flake is marked in red and the line
where the reference beam $\mathcal E_{\rm ref}$ crosses the bottom edge of the top hBN
flake in white.}
    \label{fig:fig1}
\end{figure}

A microscopy image of our device is presented in
Fig.~\ref{fig:fig1}(a), where the MoSe$_2$ monolayer flake (marked
in red) is encapsulated between thin bottom (dashed white) and top
(solid white) hBN films. The bottom graphite flake (dashed black)
lies under the hBN spacer, while the top few-layer graphene flakes
(solid black) are adjacent to the MoSe$_2$ as shown in the layer schematic in the top right inset. The graphite and
graphene operate as bottom and top gates, respectively. Details
regarding the fabrication of our device are provided in the
Supporting Information (SI). We have fabricated and investigated two samples of the same design, which show very similar results throughout the studied properties.

The device permits us to control the free electron density $n_{\rm
e}$ by applying a gate bias $U$ via the gold bands, as shown in
yellow in Fig.~\ref{fig:fig1}(a), connecting with both graphite and
graphene layers acting as electrodes. When increasing the bias and therefore $n_{\rm e}$ we
observe modifications of the optical response of the monolayer, both
in the transition energy and the spectral line shapes, as shown in
the white-light reflection contrast spectra presented in
Fig.~\ref{fig:fig1}(b). We note that by varying $n_{\rm e}$ we can
alter the relative spectral intensities of X and the X$^-$, as
quantified in (c). We see that already the linear optical response,
as observed in reflectance, is tuned when increasing $n_{\rm e}$, as
the oscillator strength shifts from X to X$^-$ -- we will come back
to this point in the last section.

To infer the coherent nonlinear response of excitons, we perform
FWM microscopy, similarly as in our recent
works~\cite{JakubczykNanoLett16}. We use three, co-linearly polarized
pulses of 150 femtosecond duration, here labeled as
$\ensuremath{{\cal E}_{1,2,3}}$, that can be tuned into resonance
with the excitonic transitions. The beams are focused down to a
diffraction-limited spot on the sample's surface, placed in a helium
flow cryostat, setting the temperature at $T=8$\,K for all
experiments. The complex-valued FWM signal, generated in a standard,
so-called photon echo, configuration is retrieved in reflectance by
combining optical heterodyning and spectral interferometry with a
reference pulse~\cite{LangbeinOL06}. The reference is focussed through
the same objective (${\rm NA}=0.65$) below $\ensuremath{{\cal E}_{1,2,3}}$
and is reflected from the neighboring hBN without MoSe$_2$, as shown
in Fig.~\ref{fig:fig1}(a) by the white dots.

To investigate the exciton's coherence dynamics we measure the FWM
amplitude as a function of time delay $\tau_{12}$ between the first
and the second arriving pulses, while fixing $\tau_{23}$.
Conversely, the exciton's population dynamics is probed in FWM when
varying the delay $\tau_{23}$, while fixing $\tau_{12}$. These pulse
sequences are depicted in Fig.~\ref{fig:fig2}(a) and
Fig.~\ref{fig:fig3}(a), respectively. We note that for all
experiments we keep the same excitation conditions of $0.3$\,\textmu
W average power for each of the beams, generating an exciton density
of a few $10^{10}$cm$^{-2}$~\cite{JakubczykNanoLett16,BoulePRM20}. To
confirm the expected linear scalings of the FWM amplitude with the
pulse areas of $\ensuremath{{\cal E}_{1,2,3}}$ for small powers, we
checked the FWM power dependence (see SI Fig. S1), also monitoring
that the excitation induced dephasing can be neglected for this
range of exciton densities~\cite{BoulePRM20}.

To characterize the overall device, we set $U=-0.5\,$V and perform a
FWM amplitude mapping across the MoSe$_2$ flake for
$\tau_{12}=1.2\,$ps, by scanning the objective's position. The
resulting time-integrated FWM image is presented in
Fig.~\ref{fig:fig1}(d) (for detailed analysis of this imaging see
SI Figs S2, S3, and S4). The dominant signals reflect the
third-order optical susceptibility of the MoSe$_2$ monolayer (red
line). Note, that the generated FWM is weaker in the upper part of
the flake, which is due to the reduced reflectance of the reference
pulse $\mathcal E_{\rm ref}$ when impinging on the top hBN. This crossover between low and
high FWM amplitude regions (marked in white) faithfully follows the
shape of the top hBN's edge when $\mathcal E_{\rm ref}$ leaves the top hBN, while the other pulses $\mathcal E_{1,2,3}$ still probe the fully encapsulated MoSe$_2$ monolayer.

In time-resolved FWM, a typical case presented in
Fig.~\ref{fig:fig2}(b), without introducing free carriers we observe
a more or less pronounced photon
echo~\cite{MoodyNatCom15,JakubczykNanoLett16,Jakubczyk2DMat17}, which
is a consequence of inhomogeneous broadening $\hbar\sigma$. The latter
induces rephasing of the signal for $t=\tau_{12}$, such that
normally the echo is aligned with the diagonal in the
$(t,\tau_{12})$-space, as depicted as dashed diagonal in (b).
However, for short delays $\tau_{12}<0.5\,$ps we spot deviations from
the typical echo. In this range the signal is retarded in time,
such that the transient is not following the diagonal, and is
instead curved, into a comet-like shape. The echo distortion is
readily reproduced, when taking into account the local field
effect~\cite{HahnNJP21}, representing an effective exciton-exciton
interaction, and is described in leading order in the local field
coupling by
\begin{eqnarray}\label{Eq:echo}
|p_{\rm FWM}(t,\,\tau_{12})| &\sim& \Theta(\tau_{12})\Theta(t) t e^{-\beta(t+\tau_{12}) - \frac 12 \sigma^2(t-\tau_{12})^2} \nonumber\\
&+& \Theta(-\tau_{12})\Theta(t+\tau_{12}) (t+\tau_{12})e^{-\beta(t-\tau_{12})-\frac 12\sigma^2(t-\tau_{12})^2}\,.
\end{eqnarray}
The fitted $(t,\,\tau_{12})$-dynamics is presented in
Fig.~\ref{fig:fig2}(c), from which we extract homogeneous $\beta$
and inhomogeneous $\sigma$ dephasing rates and corresponding
full-width at half-maxima (FWHM) spectral broadenings
$\hbar\beta=0.86\,$meV and
$2\sqrt{2\ln(2)}\hbar\sigma=\hbar\tilde{\sigma}=3.9\,$meV,
respectively. By inspecting spectrally-resolved FWM mappings, we
reveal typical fluctuations of the exciton's central energy
($\approx \pm5\,$meV) and line widths ($\approx \pm1\,$meV), as
shown in Fig.~S2 in the SI, attributed to varying strain generated
during the sample
assembly~\cite{NiehuesNanoLett18,Khatibi2DMat19,RajaNatureNano19}. We
therefore fix the excitation spot for the entire experiment, as
marked in Fig.~\ref{fig:fig1}(a), to probe an area of relatively
small $\sigma$ within the distribution. The same mapping is
performed for the time-resolved FWM signal, which allows to extract
homogeneous and inhomogeneous broadenings, summarized in SI Fig.~S3.
We find an expected and pronounced correlation between $\sigma$ and
the FWHM of the FWM spectra, as shown in SI Fig.~S4. There, we also
find a correlation between the exciton transition energy and the
inhomogeneous broadening.

\section{Coherence dynamics of neutral \& charged excitons}

\begin{figure}
    \centering
    \includegraphics[width=0.6\columnwidth]{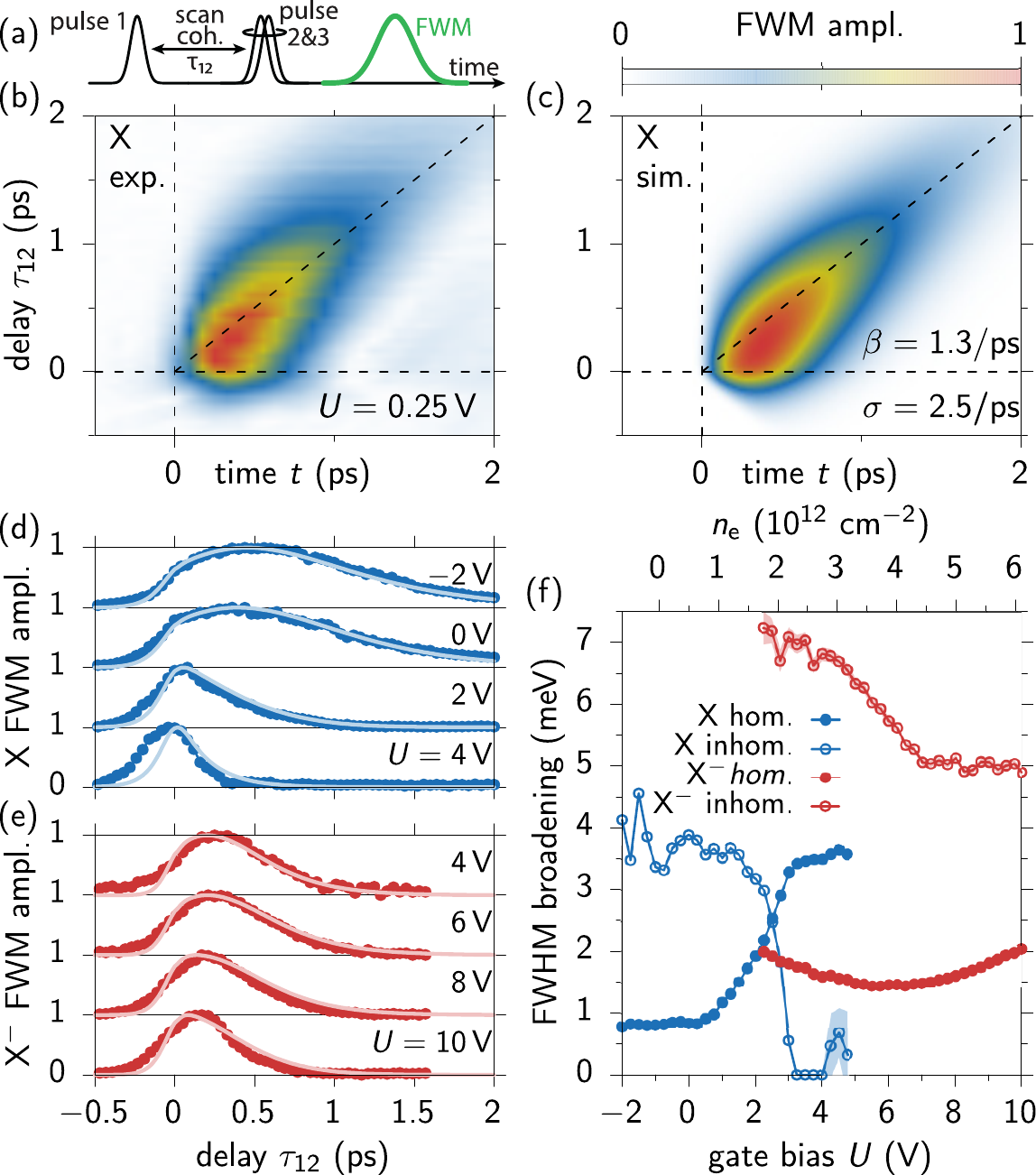}
    \caption{\textbf{X and X$^-$ dephasing versus applied
gate bias.} (a) Scheme of the three-pulse FWM sequence probing
coherence dynamics via $\tau_{12}$. (b) Measured time-resolved FWM
amplitude showing a clear photon echo. (c) Theoretical fit of (b)
employing the local field model with the fitted homogeneous and
inhomogeneous dephasing rates $\beta$ and $\sigma$ as given in the
plot. (d, e) Exemplary time-integrated FWM amplitude dynamics as a
function of $\tau_{12}$ for different gates biases. (d) For X and
(e) for X$^-$. (f) Homogeneous (filled circles) and inhomogeneous
(open circles) FWHM line widths extracted from (d, e) and
time-resolved data as a function of gates bias with X in blue and
X$^-$ in red. The shaded areas show the uncertainties.}
    \label{fig:fig2}
\end{figure}

We now proceed to the investigation of the exciton homogeneous
linewidth $\hbar\beta$ depending on the induced electron density
$n_{e}$. For this purpose, we center the excitation at the exciton
energy, which for the investigated area equals 1640\,meV, and
measure the photon echo as a function of the gate bias $U$.
Exemplary exciton coherence dynamics, i.e., time-integrated FWM
amplitudes as a function of $\tau_{12}$, are shown in
Fig.~\ref{fig:fig2}(d) (blue dots). Due to the local field, the
measured dynamics deviate from simple combined exponential and
Gaussian decays according to Eq.~\eref{Eq:echo}. The corresponding
fits are shown as pale blue lines and reproduce the experiments
well. The data clearly indicate a growth of the dephasing rate when
increasing the electron density (top to bottom). All fitted
homogeneous broadenings are presented in (f, blue filled circles) as
function of the gate bias, which quantifies the impression of a
strong rise of $\hbar\beta$ with increasing $n_{e}$. The fitted
dynamics also yield $\hbar\tilde{\sigma}$ as a function of $n_{e}$
(f, blue open circles). Note, that for the fits we used the entire signal dynamics in real and delay time from which we also extract the time-integrated data in Fig.~\ref{fig:fig2}(d, e). Interestingly, the inhomogeneous broadening
is suppressed efficiently, when injecting more electrons. The FWM
transient thus evolves from the photon echo to free-induction decay
like, as we explicitly show in SI Fig.~S5. This strikingly shows
that the excitons become less sensitive to the underlying disorder,
which can be static due to the strain or dynamic via fluctuating
charges. With increasing $n_{e}$, the disorder thus gets screened
more efficiently and the applied voltage can neutralize fluctuating
charges~\cite{AkbariNanoLett22}, both mechanisms making the system
less inhomogeneous. We have previously found that the reduced
inhomogeneity goes hand-in-hand with a shortening of the excitons'
radiative lifetime~\cite{JakubczykACSNano19}. The small rise of the X's inhomogeneous broadening at $U>4$~V is an artefact from the weak optical response which results in an increased uncertainty of the fit. This is also reflected by the significant increase of the error bars (shaded area).

The second resonance, occurring at 1620\,meV is the charged exciton
transition X$^-$, which we selectively address by centering the
laser pulse spectrum to this energy. We carry out the same routine as before to
determine the charged exciton's homogeneous and inhomogeneous
broadening and their dependence on $n_{\rm e}$. The results are
presented in Fig.~\ref{fig:fig2}(e) and (f) in red colors. When
comparing X's and X$^-$'s homogeneous broadenings in (f), we observe
that for the same electron density the X broadening is significantly
larger than that of X$^-$. This finding  is similar to experiments
performed on non-intentionally doped samples~\cite{Hao2DMat17}.
Following an intermediate drop with increasing $n_{\rm e}$, we
eventually also observe an increase of the charged exciton's
homogeneous broadening accompanied by a decrease in
inhomogeneous broadening (open red circles). The increased line
width of both exciton complexes is
attributed to the dephasing due to interactions with the Fermi-sea of electrons,
similarly as in past studies on semiconductor quantum
wells~\cite{SchultheisPRL86, KochPRB95}.

With this, we evidence the control the inhomogeneous broadening and
excitonic dephasings and thus the spectral line shape of the optical
transitions, via a tunable gate bias introducing free carriers into
the MoSe$_2$ monolayer. We distinguish this line broadening mechanism
from the ones previously investigated in TMDs, i.e.,
phonon-induced~\cite{MoodyNatCom15,JakubczykNanoLett16,
Jakubczyk2DMat17, ChristiansenPRL17, NiehuesNanoLett18} and
excitation-induced dephasing~\cite{JakubczykACSNano19, BoulePRM20}.

At this point, we remark that  the measured behavior of
dephasings does not match with a simple picture of the dipole (oscillator) strength
transfer from X to X$^-$ when increasing $n_{\rm e}$. Assuming that the
dephasing was entirely governed by the radiative lifetime, a
reduction for the X (an increase for X$^-$) dipole strength should
increase (reduce) the lifetime and consequently the dephasing time.
However, the opposite trend is observed in Fig.~\ref{fig:fig2}(f).
Next to the suppression of $\sigma$ via screening, the second
possible cause for this behavior is the presence of other
decay channels that increase with a growing $n_{\rm e}$, which would
result in an accelerated dephasing while remaining close to a
lifetime-limited condition. To shed light on the dominating decay
mechanism, we thus measure FWM as a function of $\tau_{23}$,
monitoring the coherent population dynamics of X and X$^-$.

\section{Population dynamics of neutral \& charged excitons}
\begin{figure}
    \centering
    \includegraphics[width=\columnwidth]{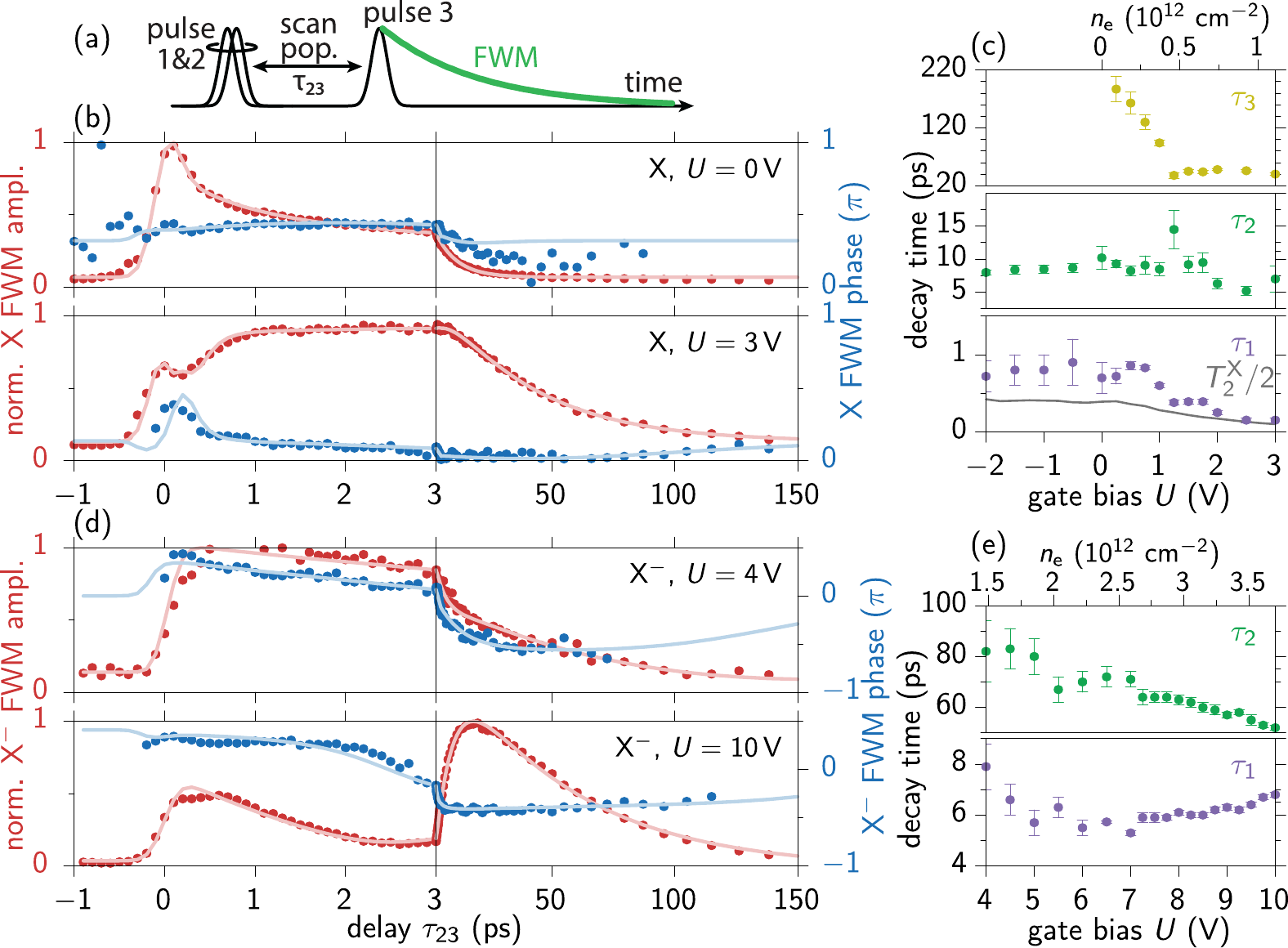}
    \caption{\textbf{X and X$^-$ population dynamics versus gate
bias.} (a) Scheme of the three-pulse FWM probing the population
dynamics. (b) Exemplary FWM amplitude (red dots) and phase (blue
dots) dynamics of X as a function of $\tau_{23}$, with fitted curves
in pale colors, according to Eq.\,1. (c) Extracted characteristic
decay times of the three identified relaxation channels as a
function of gate bias. The gray curve shows the blue data from Fig.~\ref{fig:fig2}(f) in the form $T_2^{\rm X}=1/\beta$. (d, e) Same as (b, c) but  for X$^-$.}
    \label{fig:fig3}
\end{figure}

In Fig.~\ref{fig:fig3}(b) we present the time-integrated FWM signals
of X as a function of $\tau_{23}$ for selected gate biases
(amplitude as red and phase as blue dots). As the FWM response is
measured in a coherent fashion, we can retrieve its amplitude and
phase, improving the insight into the involved decay processes
affecting the excitonic populations that occur. A natural choice to
describe this dependence is to consider a coherent superposition of
several exponential decays~\cite{ScarpelliPRB17, JakubczykACSNano19}
via

\begin{eqnarray}\label{eqn:eq1}
S_{\rm FWM}(\tau_{23},t)&= &A_{\rm off}\exp(i \varphi_{\rm off})+ \exp(i\varphi_{\rm dr}t) \Bigg\{ A_{\rm nr} \exp\left(i \varphi_{\rm nr}-\frac{\tau_{23}^2}{\tau_0^2}\right) \\
&+&\sum_n A_n \left[1+\mathrm{erf}\left(\frac{\tau_{23}}{\tau_0}-\frac{\tau_0}{2\tau_n}\right)\right] \exp \left(i\varphi_n + \frac{\tau_0^2}{2\tau_n^2}-\frac{\tau_{23}}{\tau_n}\right)\Bigg\} \nonumber
\end{eqnarray}
where $t$ is real time of the FWM transient, ($A_{\rm
nr},\varphi_{\rm nr}$) are amplitude and phase of the two-photon
absorption, ($A_n,\tau_n,\varphi_n$) are the amplitude,
characteristic time, and phase of a given decay processes,
$\varphi_{\rm dr}$ is the phase drift during the measurement, ($A_{\rm
off}$,\,$\varphi_{\rm off}$) are the amplitude and phase of the
complex offset, and $\tau_0$ is the pulse duration of the
femto-second laser of around 150\,fs.

As exemplarily represented by the fitted pale curves in
Fig.~\ref{fig:fig3}(b, top), we find that for gate bias values
$U\leq0$, which corresponds to the charge-neutrality regime, i.e.,
no free electrons, the exciton's population dynamics can be fitted
with a bi-exponential decay with characteristic timescales
$\tau_1<1\,$ps and $\tau_2\approx 7\,$ps. We plot the extracted
timescales of the relaxation processes in Fig.~\ref{fig:fig3}(c).
This result is in agreement with previous
studies~\cite{JakubczykNanoLett16,ScarpelliPRB17}, with the faster
component being attributed to the exciton's decay, which can be due
to several mechanisms: radiative recombination, non-radiative
scattering into various momentum-dark states and localized states
generated by the disorder, and finally X to X$^-$ conversion. The
longer decay stems from the relaxation of thermalized higher
$k$-vector exciton states, which scatter back into the light-cone
(i.e. states with momentum values $|k|<n\omega/c$) through
non-radiative processes and subsequently contribute to the FWM
signal.

With increasing gate bias $U$ and thus $n_{\rm e}$, the description
of the observed population dynamics requires the introduction of a
third, slower relaxation process $\tau_3$. Earlier FWM experiments
on samples naturally doped with electrons also indicate the presence
of this long-lived component and attribute it to a decaying
population of excitons with spin-forbidden
transitions~\cite{ScarpelliPRB17}. As the disorder is
screened more efficiently with increasing $n_{\rm e}$, we
tentatively suggest that the observed faster dynamics could be also
due to a weaker exciton localization.

The most pronounced change in the population dynamics, caused by the
increase of $n_{\rm e}$, can be observed during the first few
picoseconds in Fig.~\ref{fig:fig3}(a). For positive $U$ in
Fig.~\ref{fig:fig3}(c), we observe a rapid increase of the fastest
relaxation rate. This behaviour, fortified by the shift of the FWM
phase, occurs in the same voltage range as the previously mentioned
shortening of the exciton dephasing time (see blue dots in
Fig.~\ref{fig:fig2}(f)). Interestingly, when comparing the retrieved
$\tau_1$ values with $T^{\rm X}_2/2=1/(2\beta)$ (gray line in
Fig.~\ref{fig:fig3}(c, bottom)), we see the same trend with
increasing bias. In particular, we find that for large gate biases
the measurements approach $\tau_1=T^{\rm X}_2/2$ for $U\approx 3\,$V
and therefore the special situation of lifetime-limited dephasing.
This result is also in line with the previous finding that the
exciton looses nearly all inhomogeneous broadening for $U>3\,$V and
therefore the dominant dephasing mechanism for smaller $U$.

In Fig.~\ref{fig:fig3}(d, e) we present analogous data from the
measurements where the excitation energy is tuned to the charged
exciton transition X$^-$. In this case, we only observe two distinct
relaxation processes~\cite{ScarpelliPRB17, ZipfelPRB22} contributing
to the investigated range of $\tau_{23}$. X$^-$ is the lowest energy
state and -- in contrast to X -- cannot relax to other excitonic
states. Consequently, the initial decay of X$^-$ of around 6\,ps,
reflects its radiative and non-radiative recombination. We note here
also, that the simple extension of the light-cone escape process as
for the neutral excitons is now not valid for X$^-$. The free
carrier leftover after the recombination allows for the additional
momentum transfer and recombination of X$^-$ from non-zero
k-vectors~\cite{CiulinPRB00, KossackiJoPCM03}.

\begin{figure}
    \centering
    \includegraphics[width=\columnwidth]{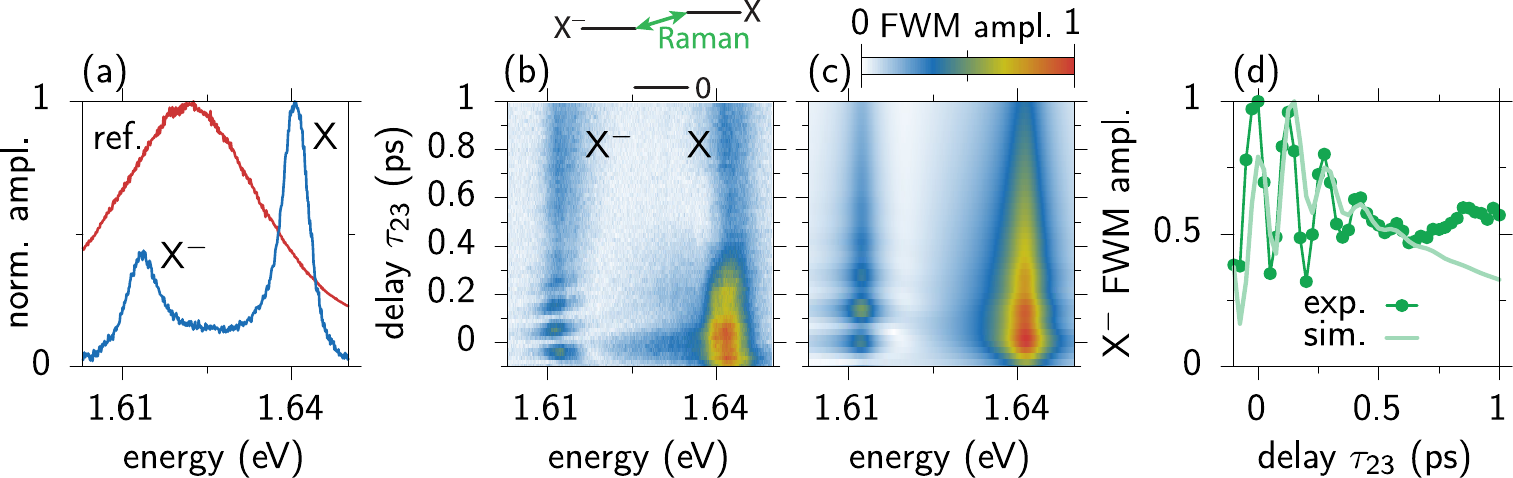}
    \caption{\textbf{Quantum beat of the X-X$^{\boldsymbol -}$ Raman coherence revealed in the population dynamics.} (a) Amplitude of the FWM spectra of X and X$^-$ in blue together with the amplitude of reference spectrum in red. (b) Spectral dynamics of the FWM amplitude as a function of $\tau_{23}$ showing a clear beating of X$^-$. (c) Theoretical simulation of (b). (d) Dynamics of the X$^-$ single after spectral integration over the respective peak. Experiment as green dots and simulation as pale line.}
    \label{fig:fig4}
\end{figure}

\section{Controlled coherent coupling between neutral \& charged excitons}
So far, we have investigated the coherence dynamics of X and X$^-$
separately, by selectively addressing the respective resonances.
Conversely, when the two complexes are excited simultaneously, both
exciton species coexist and interact with each other. To trigger the
interplay between X and X$^-$ and reveal their coupling, we excite
them in tandem, as shown spectrally in Fig.~\ref{fig:fig4}(a) by the
laser spectrum in red and the FWM spectrum in blue. The measured
spectrally resolved population dynamics are presented in (b).
Interestingly, for the initial delays $\tau_{23}< 0.5\,$ps, the FWM
displays a beating particularly pronounced on the X$^-$ resonances.
The first driving beam induces the Raman coherence between X and
X$^-$ as depicted schematically in (b). Due to the spectral
splitting $\delta$ between X and X$^-$, the density grating
generated by the second beam oscillates with the period
$2\pi\hbar/\delta\approx0.14\,$ps. The FWM released by the probe
therefore shows the Raman quantum beat~\cite{MermillodOptica16}, as
confirmed by the simulation in (c), and directly shown in (d) as
time traces after spectrally integrating over the X$^-$ peak. This
result indicates that X and X$^-$ are Raman-coupled, as they share a
common ground state. Note, that the simulation (pale green line)
only considers a single exciton decay channel, while in the
experiment (green dots) the interplay between different relaxation
paths leads to the rising signal for $\tau_{23}>0.5\,$ps (see
Fig.~\ref{fig:fig3}).

\begin{figure}
    \centering
    \includegraphics[width=1.\columnwidth]{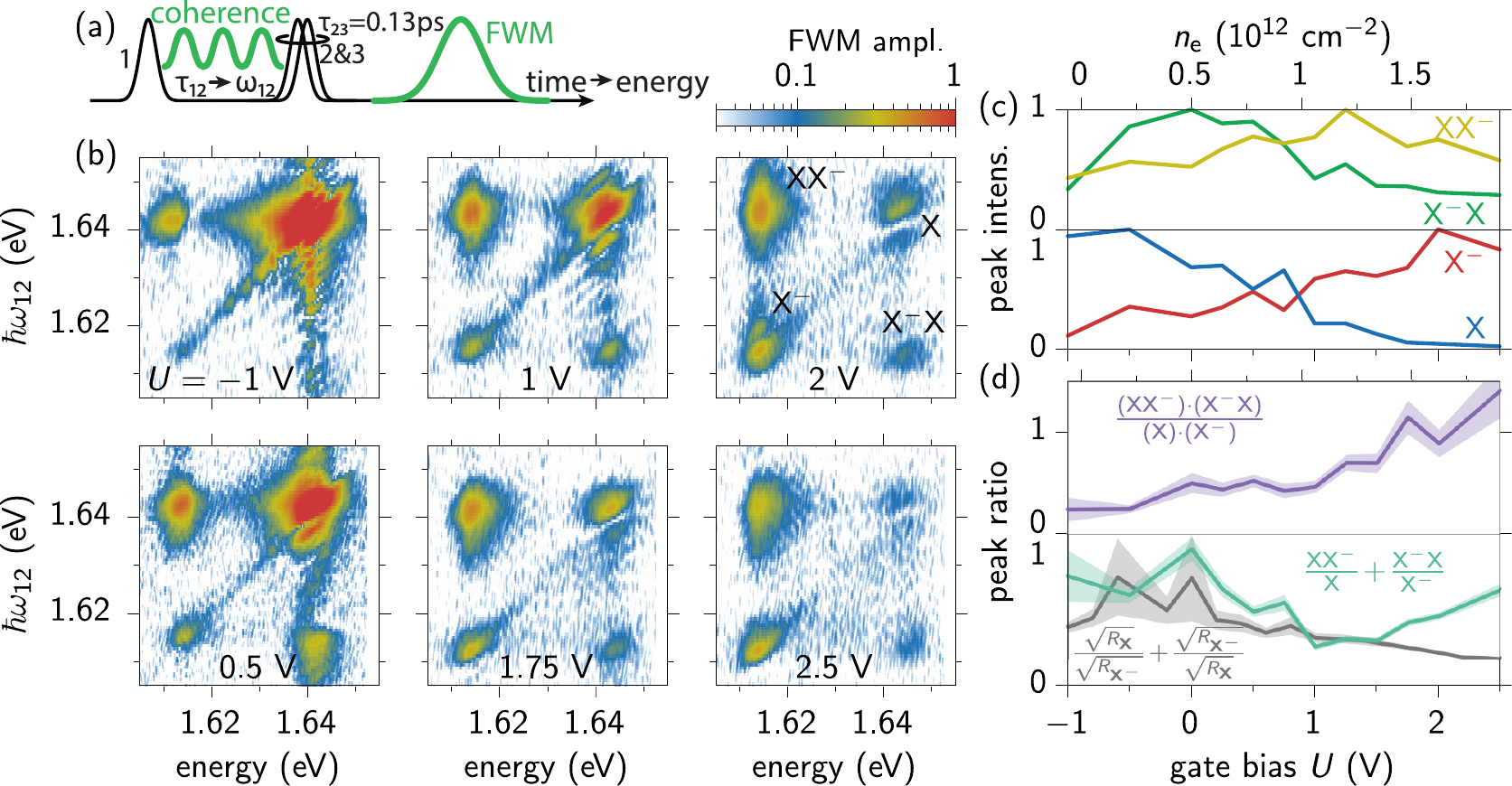}
    \caption{\textbf{Phase-referenced two-dimensional FWM spectroscopy of X and X$^{\boldsymbol -}$.} (a)\,The pulse sequence
used in the 2D FWM experiment. $\tau_{12}$ is scanned, while the
other delay is set to $\tau_{23}=0.13\,$ps maximizing the coherent
coupling. (b) Examples of 2D FWM spectra for different gate biases
as given in the plots. (c) Integrated peak amplitudes of the four
peak X, X, XX$^-$, and X$^-$X (marked in (b)) as function of gate
bias. (d) Characteristic peak rations extracted from (c) as dots and
from Fig.~\ref{fig:fig1}(c) as gray line. The shaded areas in (c, d) mark the uncertainty ranges.}
    \label{fig:fig5}
\end{figure}

A beating of a similar origin is observed in the coherence dynamics,
probed by the $\tau_{12}$-dependence, as shown in SI
Fig.~S6. When coherently coupled, the first-order absorptions of X
and X$^-$, created by the leading pulse, are mutually converted into
the FWM of X$^-$ and X, respectively, by the following two pulses,
as schematically shown in Fig.~\ref{fig:fig5}(a). To pinpoint this
phenomenon we perform two-dimensional (2D) FWM
spectroscopy~\cite{LiPRL06, MoodyPRL14}. In this approach,
originating from nuclear magnetic resonance spectroscopy, Fourier
transformations are performed along two time axes: the direct time axis and
the indirect delay axis~\cite{HammCambridge11,
MoodyAPX17,SmallwoodLPR18}. In our case, the transform along the
direct axis is automatically performed by the optical spectrometer,
yielding the FWM spectra. Conversely, the Fourier transform of the
indirect axis has to be recovered from the $\tau_{12}$-sequence. To
track the FWM phase when varying $\tau_{12}$, we apply the
phase-referencing scheme~\cite{DelmontePRB17}, which overcomes the
need for an active phase-stabilization and permits us to accurately
perform the Fourier transform along $\tau_{12}$, yielding the energy
axis $\hbar\omega_{12}$.

A conclusive display for the coherent coupling between X and X$^-$
is presented in Fig.~\ref{fig:fig5}\,(b), showing 2D FWM amplitude
spectra for selected gate voltages as labeled in the pots. The delay
between pulses 2 and 3 was chosen to
$\tau_{23}=0.13\,$ps, which locates us in the coherent coupling
range as demonstrated in the SI Fig.~S7. We have
checked that for $\tau_{23}>1\,$ps the X and X$^-$ coupling
is present (see Fig.~S8), although dominated by an incoherent
population transfer between the two complexes~\cite{MoodyNatCom15}.
Conceptually, the 2D FWM spectra of our V-shaped system consist of only four peaks.
For the diagonal pair, labeled X$^-$ and X, detection energy and
$\hbar\omega_{12}$ energy are identical: FWM emission arises from the same
absorption. Decisively, we clearly detect off-diagonals, labeled
XX$^-$ and X$^-$X. This means that the FWM of the charged exciton is
also driven by the neutral exciton's first-order absorption, and
vice-versa, respectively. Such coherent coupling was previously reported in
semiconductor nanostructures~\cite{LangbeinOL06, KasprzakNPho11,
MoodyPRL14}, including TMDs~\cite{SinghPRL14,HaoNanoLett16,
HaoNatCom17}. Here, thanks to the tunability of $n_{\rm e}$ in this
gated MoSe$_2$, we obviously find that the coherent nonlinear
optical response of the X\,-\,X$^-$ complex can be controlled simply
by applying an external bias.

The variation of the peak amplitudes in the 2D spectra with changing
gate bias has two potential reasons: (i) Change of the dipole
strengths of X and X$^-$ due to additional free charges, as already
demonstrated in the reflectivity measurement in
Fig.~\ref{fig:fig1}(b), (ii) The coherence transfer between both
exciton species is affected by the free carriers. To disentangle
these two effects we need to quantify the strength of the coherent
coupling depending on $n_{\rm e}$. In panel (c) we plot the peaks'
integrated amplitudes versus the gate bias, where we expectedly find
that X (blue) clearly drops, while X$^-$ (red) builds up when
increasing the carrier density. At the same time the off-diagonal
peaks (green and yellow) -- representing the coupling -- show no
clear trend. As mentioned before, the variation of the reflectivity
spectrum upon an applied gate bias indicates a change of the dipole
strengths of X and X$^-$ and therefore of the pulse areas
$\theta_{\rm X}$ and $\theta_{{\rm X}^-}$ applied to the two
transitions. Note, that we do not consider any specific origin for
the change of the strength of optical response of X and X$^-$. Now,
in order to extract the strength of the X\,-\,X$^-$ coupling found
in the 2D spectra, we can extract a quantity that does not depend on
the pulse areas. To achieve this, we note that the amplitudes of
diagonal peaks scale as $\theta_{\rm X}^3$ and $\theta_{{\rm
X}-}^3$, whereas the off-diagonals follow the scaling $\theta_{\rm
X}^2\theta_{{\rm X}-}^{}$ for XX$^-$ and $\theta_{\rm
X}^{}\theta_{{\rm X}-}^2$ for X$^-$X. Plotting now the expression,
XX$^-\cdot$X$^-$X/(X$\cdot$X$^-$) in Fig.~\ref{fig:fig5}(d, top) the
pulse areas and therefore the dipole strengths cancel out. We see
though, that this quantity increases with the electron density,
indicating the increase of the coupling strength.

Another insightful quantity that can be extracted from the
reflectivity spectra and the 2D spectra is the sum of peak ratios.
For the 2D spectra we calculate XX$^-$/X+X$^-$X/X$^-$, which
translates into pulse areas as $\theta_{{\rm X}^-}/\theta_{\rm
X}+\theta_{\rm X}/\theta_{{\rm X}^-}$. To extract an equivalent
quantity from the reflectivity spectra we need to calculate
$\sqrt{R_{{\rm X}^-}}/\sqrt{R_{\rm X}}+\sqrt{R_{\rm
X}}/\sqrt{R_{{\rm X}^-}}$ from the peak intensities $R_{\rm X}$ and
$R_{{\rm X}^-}$. The two quantities are plotted (normalized to
unity) in Fig.~\ref{fig:fig5}(d) as turquoise dots for 2D and gray
line for reflectivity. The behaviour of both curves is rather
unspecific for negative and small gate voltages, where X$^-$ is only
weakly addressed if present at all. Note, that in this range of bias values the X$^-$ is very weak which is leading to the relatively large uncertainties (shaded areas). Therefore, the peak ratios in this range are not particularly significant and we do not expect to draw conclusions from this bias range. However, the 2D spectra
result shows a clearly growing trend for increasing charge
densities, while the reflectivity results shrink. This discrepancy
is a second hint that the variation of the dipole strengths with
increasing gate bias alone, cannot fully explain the variation of
the coherent coupling between X and X$^-$ observed in 2D FWM.

\section{Conclusions \& Outlook}
FWM spectroscopy methods are powerful tools that have led in the recent years to a significant improvement of the understanding of the rich exciton physics in TMDs. This has established the method as a versatile technique for studies of atomically thin materials and TMD-based heterostructures as detailed in the Introduction. However, so far a particular focus on the important issue of the free carrier influence on the studied phenomena has been missing. A factor that fundamentally affects the optical response of TMDs, and governs even the most basic characteristics like the number and respective intensities of different exciton states.

Using FWM micro-spectroscopy, we have shown that the homogeneous
linewidth and population lifetime of excitonic complexes hosted by a
MoSe$_2$--monolayer-based van der Waals heterostructure can be tuned
via the free electron density, which is injected by a gate bias
applied to our device. With increasing gate bias, the exciton's
inhomogeneous broadening decreases, reflecting the screening of the
disorder via free the electron gas, which also increases the
excitons' radiative decay rates. Conversely, the homogeneous
broadening increases, which is attributed to the combined increase
of the radiative decay rate and the conversion rate of the neutral
exciton towards the charged one. By exciting the neutral (X) and
charged exciton (X$^-$) simultaneously and probing the population
dynamics we demonstrated that the coherence between X and X$^-$
leads to a characteristic quantum beat in the FWM signal. By then
performing two-dimensional FWM spectroscopy for a variety of applied
gate biases, we have further demonstrated that the X\,-\,X$^-$
coherent coupling can be controlled by the gate voltage, and hence
by the free electron density. Through considering specific peak
ratios, we were able to demonstrate that the change of coherent
coupling can be disentangled from the variation of dipole strengths
arising from the injection of free carries. An increase of the
coupling strength with $n_{\rm e}$ could be linked with the
screening of disorder via the electron gas. This illustrates the
utility and versatility of ultrafast nonlinear spectroscopy in
investigating optical responses of excitonic systems, going beyond
the capabilities of linear methods.

In future developments, by performing FWM with spatially separated driving
beams, while using devices hosting highly diffusive excitons, it will be possible to achieve non-local coherent coupling in a two-dimensional
semiconductor. Combining this approach with two-color FWM spectroscopy would
also permit to selectively address the coherence transfer between
neutral and charged excitons. Our findings yield exciting prospects
for forthcoming investigations of coherent phenomena in the context
of recent discoveries of strongly-correlated exciton phases in
solids~\cite{SmolenskiNature21}, optically probed quantum Hall
states~\cite{PopertNanoLett22}, moir{\'e}
superlattices~\cite{TranNature19}, and magnetic two-dimensional
materials~\cite{WilsonNatMat21}. In practice, coherent nonlinear
spectroscopy could be used to optically infer the dephasing
processes of the Umklapp branches of TMD Wigner crystals and
fractional quantum Hall states in graphene.

\ack
This work was supported by the Polish National Science Centre under
decisions DEC-2020/39/B/ST3/03251. The Warsaw team (A.R., P.K. and
M.P.) acknowledges support from the ATOMOPTO project (TEAM program
of the Foundation for Polish Science, co-financed by the EU within
the ERDFund), CNRS via IRP 2D Materials, EU Graphene Flagship. M.P acknowledges support by the Foundation for Polish Science (MAB/2018/9 Grant within the IRA Program financed by EU within SG OP Program). The
Polish participation in the European Magnetic Field Laboratory
(EMFL) is supported by the DIR/WK/2018/07 of MEiN of Poland. A. R.
acknowledges support of this work from the Diamentowy Grant under
decision DI2017008347 of MEiN of Poland. K.~W. and T. T. acknowledge
support from the Elemental Strategy Initiative conducted by the
MEXT, Japan, (grant no. JPMXP0112101001), JSPS KAKENHI (grant no.
JP20H00354), and the CREST (JPMJCR15F3), JST. J. H. acknowledges
support from EPSRC doctoral prize fellowship. D. W. was supported by the Science Foundation Ireland (SFI) under Grant 18/RP/6236. We thank Ronald Cox
and Julien Renard for their constructive comments on the manuscript, Rafa\l{} Bo\.zek for the AFM measurement of the hBN thicknesses, and Tilmann Kuhn for helpful discussions.

\section*{References}
\bibliographystyle{iopart-num}

\providecommand{\newblock}{}

\end{document}


\title[]{Controlled coherent-coupling and dynamics of exciton complexes in a MoSe$_{\boldsymbol2}$ monolayer\\ {\it (Supplementary Information)}}

\author{Aleksander~Rodek$^1$, Thilo~Hahn$^2$, James~Howarth$^3$, Takashi~Taniguchi$^4$, Kenji~Watanabe$^5$, Marek~Potemski$^{1,6,7}$, Piotr~Kossacki$^1$, Daniel~Wigger$^8$, and Jacek~Kasprzak$^{1,9,10}$}

\address{$^1$Faculty of Physics, University of Warsaw, ul. Pasteura 5, 02-093 Warszawa, Poland}
\address{$^2$Institute of Solid State Theory, University of M\"unster, 48149 M\"unster, Germany}
\address{$^3$National Graphene Institute, University of Manchester, Booth St E, M13 9PL United Kingdom}
\address{$^4$International Center for Materials Nanoarchitectonics, National Institute for Materials Science,  1-1 Namiki, Tsukuba 305-0044, Japan}
\address{$^5$Research Center for Functional Materials, National Institute for Materials Science, 1-1 Namiki, Tsukuba 305-0044, Japan}
\address{$^6$Laboratoire National des Champs Magn\'{e}tiques Intenses, CNRS-UGA-UPS-INSA-EMFL, 25 Av. des Martyrs, 38042 Grenoble, France}
\address{$^7$CENTERA Labs, Institute of High Pressure Physics, PAS, 01- 142 Warsaw, Poland}
\address{$^8$School of Physics, Trinity College Dublin, Dublin 2, Ireland}
\address{$^9$Univ. Grenoble Alpes, CNRS, Grenoble INP, Institut N\'{e}el, 38000 Grenoble, France}
\address{$^{10}$Walter Schottky Institut and TUM School of Natural Sciences, Technische Universit\"at M\"unchen, 85748 Garching, Germany}

\ead{aleksander.rodek@fuw.edu.pl, piotr.kossacki@fuw.edu.pl, jacek.kasprzak@neel.cnrs.fr}

\vspace{10pt}
\begin{indented}
\item[]January 2023
\end{indented}

\begin{abstract}~\\
This document contains:\\
S1 Sample Fabrication\\
S2 FWM pulse power dependence\\
S3 Spatial characterization of the FWM response\\
S4 Reduction of inhomogeneous broadening with gate bias\\
S5 Coherence beat between X and X$^{ -}$\\
S6 2D peak amplitudes for varying ${\tau_{23}}$\\
S7 2D FWM spectrum in the incoherent coupling range
\end{abstract}

%
%
%
%
%

\newpage

\section{Sample Fabrication}
The MoSe$_2$ crystals are obtained from HQ graphene. The monolayer
was extracted from a bulk material via exfoliation. The hBN crystals
were provided by the partners from NIMS. The MoSe$_2$ was
encapsulated between bottom and top hBN thin films of respective
thicknesses 52\,nm and $\sim$5\,nm, as determined by AFM
measurements. For the back gate, graphene/graphite was exfoliated
onto 290\,nm SiO$_2$/Si substrate, following an O$_2$/Ar plasma etch
in order to improve crystal adhesion to the surface.
Heterostructures were assembled by consecutive dry-peel transfers of
hBN, MoSe$_2$ and graphene onto the exfoliated back
gate~\cite{KretininNanoLett14, WithersNatMat15}. We employed the
PMMA membrane technique for sample fabrication, exfoliating directly
onto substrates coated with a bilayer of PMMA/PMGI polymers. While
the crystals were prepared in air, all transfers were conducted in
inert, Argon filled glove-box environment.

Electrical contacts were patterned by electron-beam lithography
followed by evaporation of 3\,nm of Cr followed by 50\,nm of Au. To
tune n$_{\rm e}$, we vary the gate bias, supplied by the
external voltage source linked to the structure via the BNC
feed-through connectors of the cryostat. In this work we use a pair
of twin devices, A and B. The discussed results were obtained on the
sample A, apart from those shown in Figs.~1(b, c), 4, 5, S6, and
S8, which were measured on sample B. Employing a
capacitance model~\cite{RadisavljevicNatNanotech11,MakNatMat13}, we
calibrate the free electron density in the sample A in the voltage
scan to $n_{\rm e}\approx(3.7\pm0.3)\times 10^{11} {\rm cm}^{-2}{\rm
V}^{-1}$, assuming a hBN permittivity of $\epsilon_{\rm hBN}=3.5$. The corresponding calibration performed for
the sample B yields $n_{\rm e}\approx(5.6\pm0.3)\times 10^{11} {\rm
cm}^{-2}{\rm V}^{-1}$ with the neutrality point, i.e., the absence of free electrons, at a
gate bias of $-0.5$\,V.

\section{FWM pulse power dependence}
\begin{figure}[h]
    \centering
    \includegraphics[width=0.5\columnwidth]{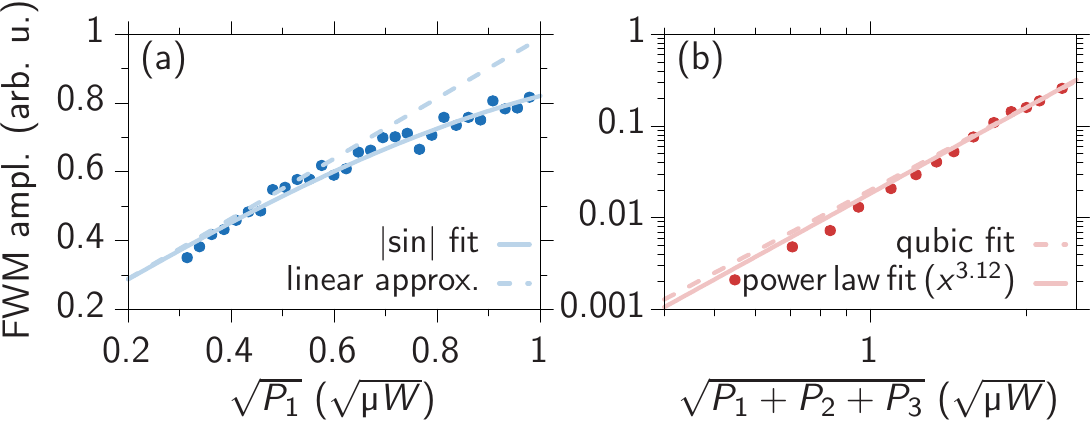}
    \caption{FWM amplitudes as a function of applied pulse powers. (a) Approximate linear scaling with the pulse amplitude of the first pulse. (b) Approximate cubic dependence of the FWM amplitude when increasing all three pulse powers.}
    \label{fig:figs1}
\end{figure}
Figure~\ref{fig:figs1} shows the pulse power dependence of the FWM amplitude. By fitting the dependence on $\sqrt{P_1}$ with a $|\sin|$ function we can confirm that our experiments were performed in the linear-response regime of the monolayer. In (b) we additionally confirm that the FWM amplitude approximately follows a cubic dependence with the third power when increasing all three laser pulse powers simultaneously.

\section{Spatial characterization of the FWM response}
\begin{figure}[h]
    \centering
    \includegraphics[width=0.5\columnwidth]{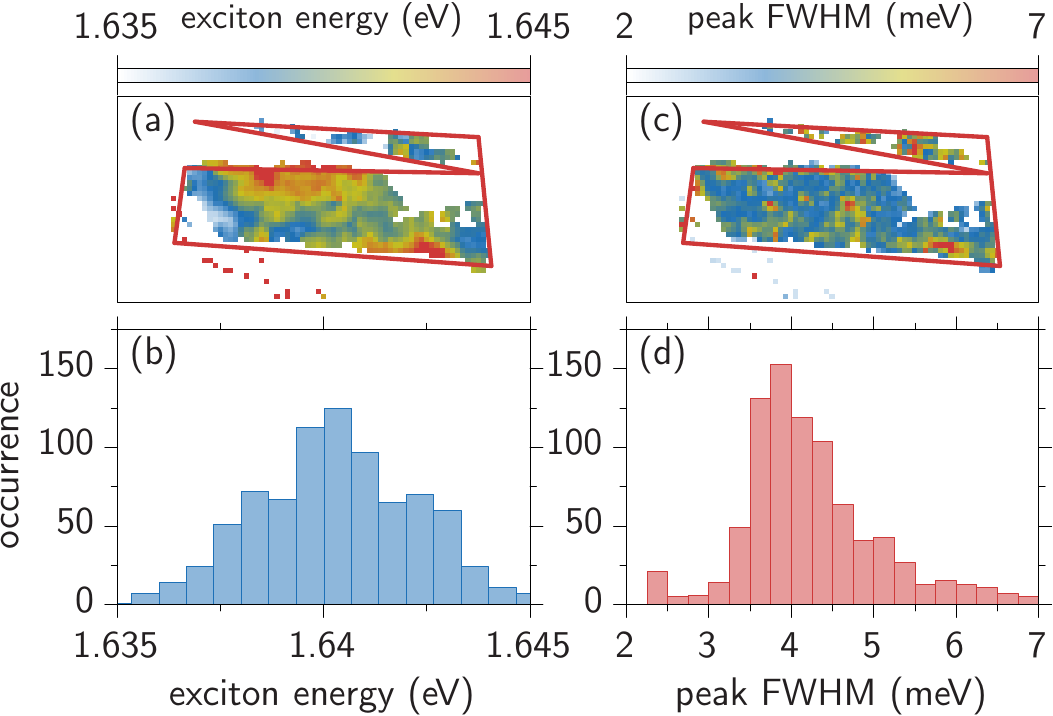}
    \caption{(a) Spatial map of the central X energy across the MoSe$_2$ sample. (b) Histogram of the data in (a). (c, d) Same as (a, b) but for the FWM peak FWHM.}
    \label{fig:figs2}
\end{figure}
Figure~\ref{fig:figs2} gathers the spectral characteristics of the FWM measurement. (a) Shows the inhomogeneous distribution of the central X energy across the monolayer and the histogram in (b) shows the respective distribution spanning over approximately 10 meV. (c) and (d) show the FWM spectral peak FWHM in the same way as (a) and (b). The spatial distribution of the peak width in (c) seems not directly correlated with the central energy in(a). The majority of the distribution in (d) of widths spans from 3 meV to 5 meV.

\begin{figure}[h]
    \centering
    \includegraphics[width=0.5\columnwidth]{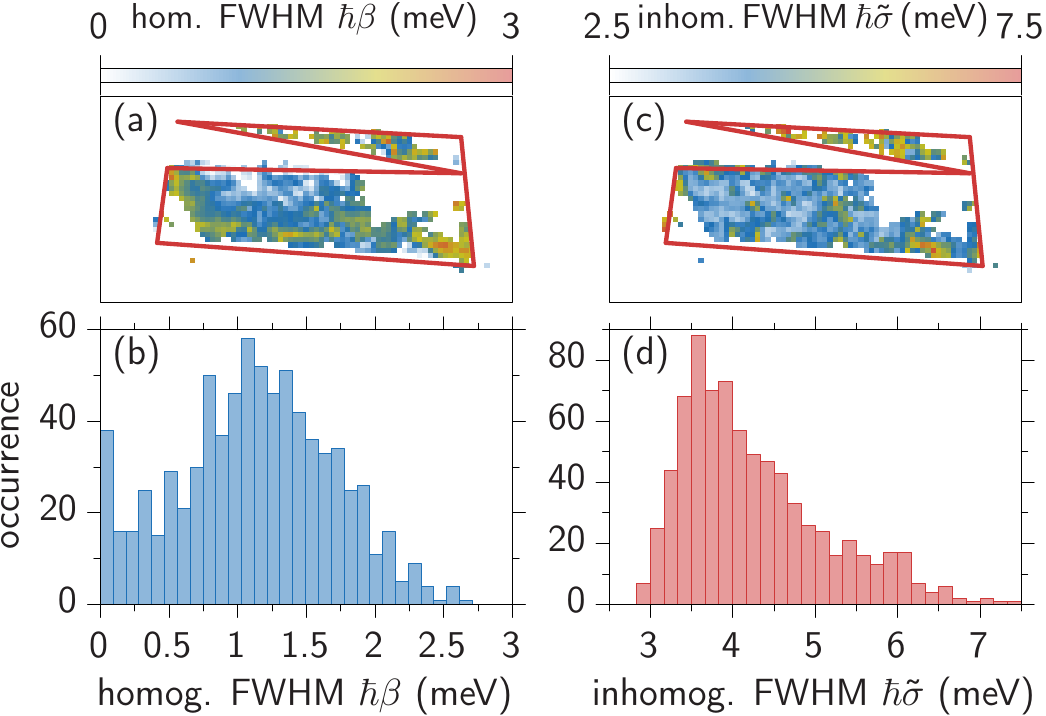}
    \caption{(a) Spatial map of the homogeneous broadening of X across the MoSe$_2$ sample. (b) Histogram of the data in (a). (c, d) Same as (a, b) but for the inhomogeneous broadening with $\tilde{\sigma}=2\sqrt{2{\rm ln}(2)}\sigma$.}
    \label{fig:figs3}
\end{figure}
Figure~\ref{fig:figs3} collects the spatial mapping of the
homogeneous (a, b) and inhomogeneous broadening (c, d) in the same
way as Fig.~\ref{fig:figs2}. The width distribution span from 0.5
meV to 2 meV for the homogeneous and from 3 meV to 6 meV for the
inhomogeneous broadening. Comparing the spatial maps in (a) and (c)
with those in Fig.~\ref{fig:figs2}(a) and (c), respectively, we see
correlations. In the regions where the X energy is larger, the
homogeneous broadening is reduced. At the same time the regions with
larger spectral widths coincide with larger inhomogeneous
broadenings. This correlation is easy to understand because the
inhomogeneous contribution is the dominating factor for the line
broadening.

\begin{figure}[h]
    \centering
    \includegraphics[width=0.5\columnwidth]{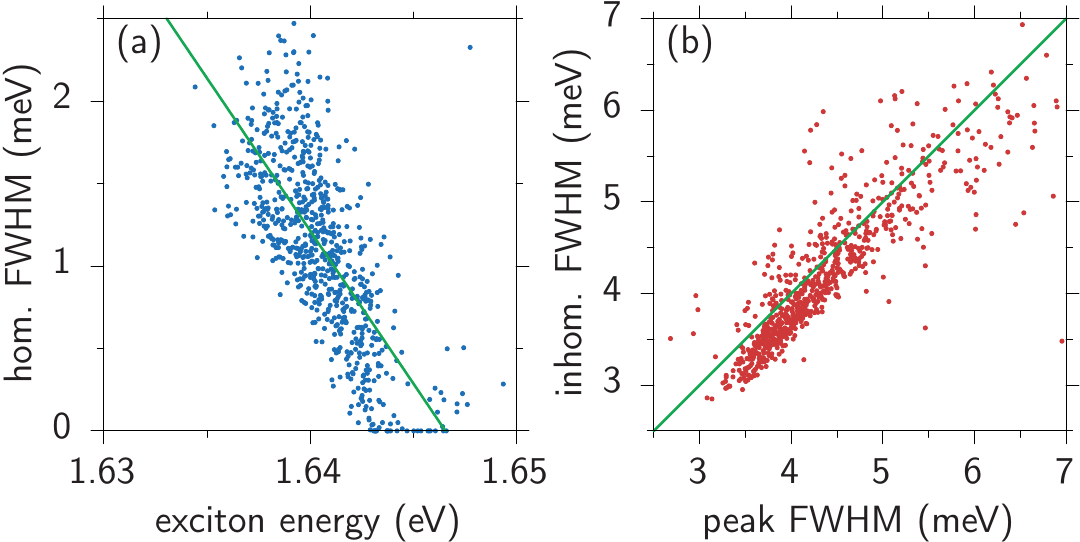}
    \caption{Correlations in the FWM measurements across the sample. (a) Comparing the exciton energy with the homogeneous broadening retrieved from the measured echos. (b) Comparing the FWM spectral width and the inhomogeneous broadening retrieved from the measured echos.}
    \label{fig:figs4}
\end{figure}
To quantitatively confirm the previously explained correlations in Fig.~\ref{fig:figs4}(a) we plot the homogeneous broadening versus the exciton energy across the monolayer and we indeed find a correlation with a negative slope. Also the correlation between inhomogeneous broadening and spectral width can be clearly confirmed in (b), where we find that the distribution of points faithfully follows the diagonal with a slight offset to larger peak widths stemming from the homogeneous contribution.

\section{Reduction of inhomogeneous broadening with gate bias}
\begin{figure}[h]
    \centering
    \includegraphics[width=0.5\columnwidth]{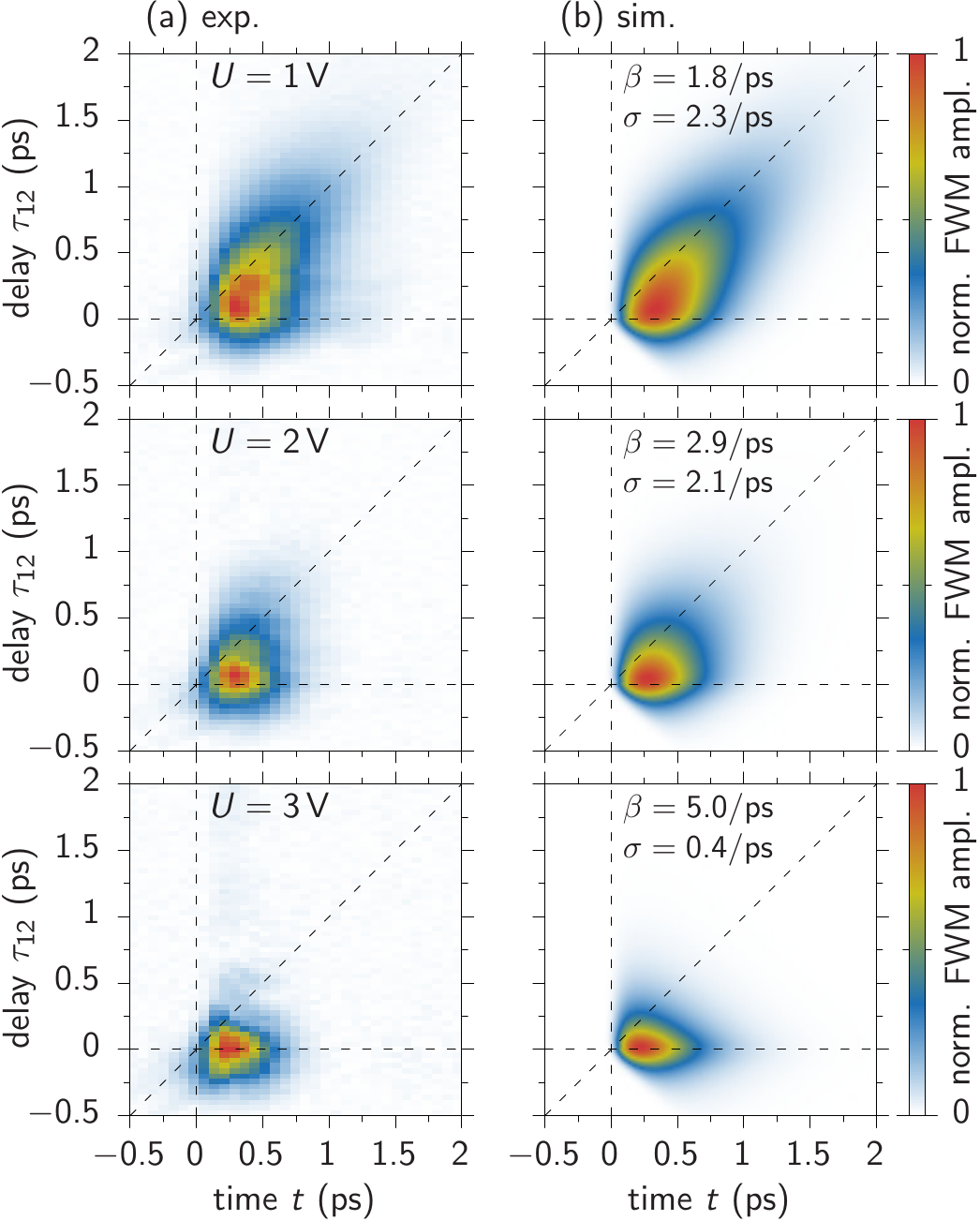}
    \caption{FWM amplitude dynamics in real time $t$ and delay $\tau_{12}$ for different gate biases increasing from top to bottom. (a) Experiment and (b) fitted simulation in the local field model. The retrieved homogeneous and inhomogeneous dephasing rates are given in the plot.}
    \label{fig:figs5}
\end{figure}
To illustrate the loss of inhomogeneous broadening with increasing gate bias $U$ in Fig.~\ref{fig:figs5}(a) plot FWM coherence dynamics. We clearly see that the initial dominant photon echo formation for $U=1\,$V (top) shrinks to a pure exponential decay for larger voltages (middle and bottom). The corresponding fits with the local field model from Eq.~(1) in the main text is shown in (b). The noted inhomogeneous dephasing rate $\sigma$ shrinks significantly from top to bottom, while the homogeneous rate $\beta$ slightly increases. The extension of the dynamics to $\tau_{12}<0$ clearly shows the requirement of the local field model to reproduce the experiments appropriately.

\section{Coherence beat between X and X$^{\boldsymbol -}$}
\begin{figure}[h]
    \centering
    \includegraphics[width=0.425\columnwidth]{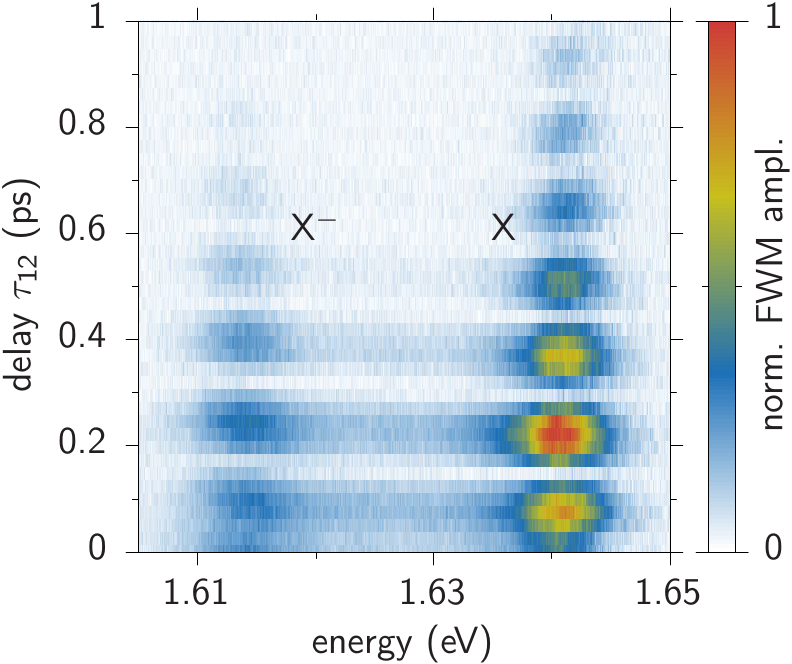}
    \caption{Normalized FWM amplitude as a function of energy and delay $\tau_{12}$ showing characteristic oscillations of X and X$^-$ that lead to the off-diagonal peaks in the 2D spectra.}
    \label{fig:figs6}
\end{figure}
For completeness in Fig.~\ref{fig:figs6} we show an exemplary FWM spectral dynamics as a function of the delay $\tau_{12}$, which represents the basis for the 2D FWM spectra in the main text. To retrieve the 2D spectra data like the one presented here are Fourier transformed with respect to the delay axis.

\section{2D peak amplitudes for varying $\boldsymbol \tau_{\boldsymbol{23}}$}
\begin{figure}[h]
    \centering
    \includegraphics[width=0.5\columnwidth]{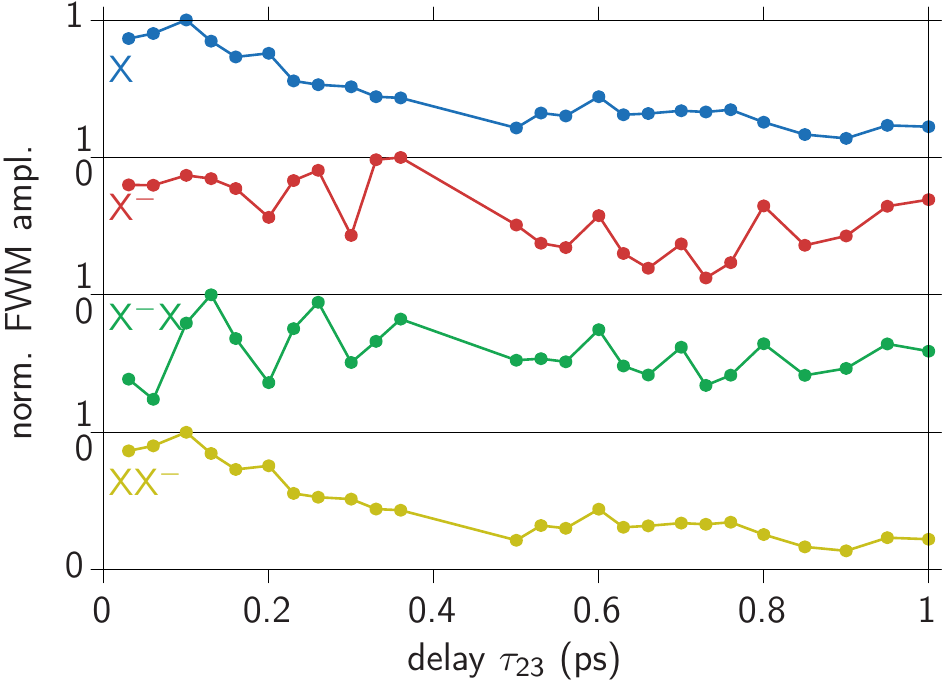}
    \caption{Integrated peak amplitudes as a function of delay $\tau_{23}$ showing clear oscillations in some of the components.}
    \label{fig:figs7}
\end{figure}
In Ref.~\cite{MoodyNatCom15} it was discussed that the appearance of
off-diagonal peaks in 2D FWM spectra retrieved from these systems
have two main contributions: (i) The coherence transfer between X
and X$^-$ and (ii) the transfer of population from X to X$^-$. By
scanning the delay $\tau_{23}$ and observing oscillatory components
in Fig.~\ref{fig:figs7} we confirm that our experiments presented in
the main text are taken in the coherence transfer range. It was
shown before that the oscillations vanish for longer delays where
the appearance of off-diagonal peaks is governed by population
transfer.

\section{2D FWM spectrum in the incoherent coupling range}
\begin{figure}[h]
    \centering
    \includegraphics[width=0.4\columnwidth]{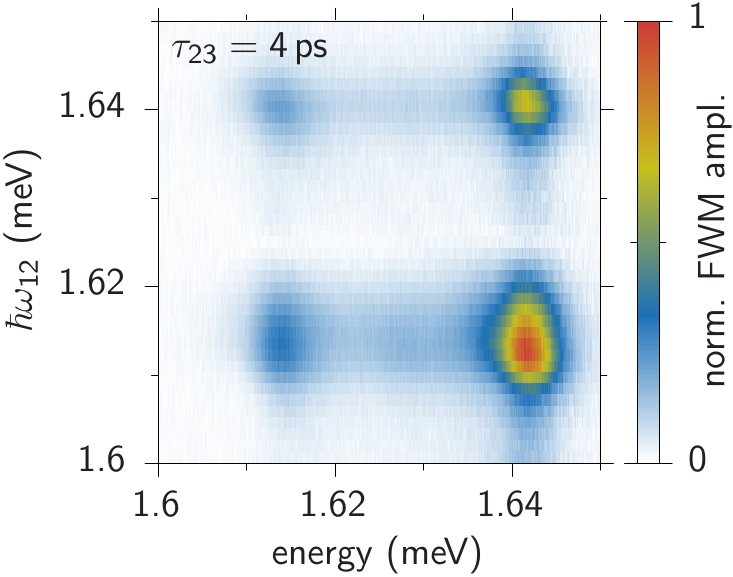}
    \caption{2D FWM spectrum for a pulse delay of $\tau_{23}=4$\,ps.}
    \label{fig:figs8}
\end{figure}
As also shown in Ref.~\cite{MoodyNatCom15}, for delay times longer than any dephasing time in the system, i.e., of the individual exciton complexes and the Raman coherence, the appearance of off-diagonal peaks in the 2D FWM spectrum arise from population transfer from X to X$^-$. An example from this regime is shown in Fig.~\ref{fig:figs8}.

\section*{References}
\bibliographystyle{iopart-num}
\providecommand{\newblock}{}